\documentclass{article}
\usepackage{makeidx}
\usepackage{graphicx}
\usepackage{amsmath} 
\usepackage{verbatim}  
\usepackage{tabularx}
\usepackage{rotating} 
\usepackage{fancyvrb}
\usepackage{float}
\usepackage[semicolon, authoryear]{natbib} 
\usepackage{url}  
\usepackage{algorithm}
\usepackage{algpseudocode} 
\usepackage{algorithmicx}
\usepackage{fancyhdr}

\usepackage{tikz}
\usetikzlibrary{calc}

\begin{document}
	\date{}

	\def\spacingset#1{\renewcommand{\baselinestretch}%
		{#1}\small\normalsize} \spacingset{1}

  	\title{\bf Significance and Replication in simple counting experiments: Distributional Null Hypothesis Testing }
		\author{Fintan Costello \\ School of Computer Science and
	Informatics,\\ University College Dublin\\
	and \\
	Paul Watts \\ Department of Theoretical Physics,\\National University of
	Ireland  Maynooth\\ }
		\maketitle

  \hfil Short title: Distributional Nulls \hfil
	
	\bigskip 
	
	\begin{abstract}
  Null Hypothesis Significance Testing (NHST) has long been  of central importance to psychology as a science, guiding theory development and underlying the application of evidence-based intervention and decision-making.   Recent years, however, have seen  growing awareness of  serious problems with NHST as it is typically used; this awareness has led to proposals to limit the use of NHST techniques,  to abandon these techniques and move to alternative statistical approaches, or even to ban the use of NHST entirely.  These proposals are premature, because the observed problems with NHST  arise as a consequence of an historically contingent, essentially unmotivated, and fundamentally incorrect, choice: that of NHST testing against point-form null hypotheses.  Using simple counting experiments we give a detailed presentation of an alternative, more general approach: that of testing against distributional nulls.  We show that this distributional approach is well-motivated mathematically, practically and experimentally, and that the use of distributional nulls addresses various problems  with the standard point-form NHST approach, avoiding issues to do with sample size and allowing a coherent estimation of the probability of replication of a given experimental result.  Rather than abandoning NHST, we should use the NHST approach in its most general form, with distributional rather than point-form null hypotheses.

	\end{abstract}

	\vfil
 	  \hfil
	\vfill
		
	\newpage
	\spacingset{1.45} 

\newcommand{\XYlabeledImage}[3]{
	\begin{tikzpicture}
	\node[inner sep=0pt] (A) {#3 };
	\node[black] (B) at ($(A.south)!-0.05!(A.north)$) {#1};
	\node[black,rotate=90] (C) at ($(A.west)!.05!(A.east)$) {#2};
	\end{tikzpicture}}
 
\newcommand{\citeg}[1]{\citeauthor{#1}'s}
\newcommand{\citen}[1]{\citeauthor{#1}}

\begin{quotation}
	{\itshape	In relation to the test of significance, we may say
		that a phenomenon is experimentally demonstrable when we know how to
		conduct an experiment which will rarely fail to give us a statistically
		significant result. \, \citep[][p. 1515]{fisher1960design}}
	
\end{quotation}

 Experimental results are useful when they demonstrate phenomena or effects in such a way as to suggest that these effects are real. This demonstration is especially important in research involving complex, interacting, and only partially understood systems, such as psychology.    Null Hypothesis Significance Testing (NHST) gives a mechanism for assessing the extent to which a given result suggests a real effect, operationalised in terms of the probability of that result arising simply by chance: if the observed result could reasonably have arisen purely by chance, the effect is probably not real.  In the NHST approach the $p$-value of a given result (the probability  of obtaining that result under the `null hypothesis' that outcomes are produced by chance processes alone) is compared against some significance criterion $\alpha$ such that results whose $p$ is less than $\alpha$ will be taken to suggest a real effect.
 
 The NHST approach has become a central part of  scientific psychology, with statistical significance   driving theoretical development and understanding, and  guiding evidence-based intervention and decision-making.  Despite this, it is becoming increasingly clear that NHST, at least as it is typically used, is fundamentally and deeply flawed.  These flaws are apparent when we consider the growing replication crisis in psychology (the finding that many statistically significant experimental results are not significant in replications), a crisis seen as `reflecting an unprecedented level of doubt among practitioners about the reliability of research findings in the field' \citep{pashler2012editors}.  These flaws are also evident in well-known effects of sample size on significance (the observation that the probability of getting a statistically significant result in the standard NHST approach increases with sample size, irrespective of the presence or absence of a true effect).  To quote \citet{thompson1998praise}: `Statistical testing becomes a tautological search for enough participants to achieve statistical significance. If we fail to reject, it is only because we've been too lazy to drag in enough participants'.  These flaws are also seen in the repeated observation that the standard NHST approach is overconfident in identifying apparently nonsensical effects (such as telepathy or precognition) as real \citep{wagenmakers2011psychologists}; to quote \citet{diaconis1991replication} `parapsychology is
 worth serious study [...because..] it offers a truly alarming massive case study of how statistics can
 mislead and be misused'.
 
 In the face of these problems there have been increasing, and increasingly widespread and vocal, calls for the wholesale abandonment of the NHST approach \citep[e.g.][]{BlakeleyAbandon2019,amrhein2018remove,hunter1997needed,carver1978case}.  We think this is premature.  Our aim in this paper is to show that all these problems with the NHST approach arise from a single source: the use of `point-form' null hypotheses in standard NHST.  This use of point-form null hypotheses is fundamentally incorrect, both mathematically and experimentally, and leads to systematically inflated judgements of statistical significance.  We argue that NHST should be based on the use of distributional-form nulls: we show that  that the use of distributional-form null hypotheses avoids all problems associated with  sample sizes in NHST,  allows for conservative rejection of the null, and allows meaningful and coherent estimation of the probability of replication of a given result.  
 
 The organisation of this paper is as follows.  In the first section we present the concept of distributional-form null hypotheses in a specific illustrative context: that of a simple `counting experiment'.  In the second section we present the normatively correct mathematical model for hypotheses in this experimental context: the Beta distribution.   In the third section we describe some properties of Beta-distributed null hypotheses. In the fourth section we describe how these properties  address the various problems arising with standard NHST.  In the fifth section we describe how the use of distributional null hypotheses allows us to effectively estimate the probability of replication of a given result.  In the sixth section we give some practical examples of using and resporting distributional null hypothesis tests, and in the final section we address some possible criticisms of these ideas of distributional NHST.

\section{Counting experiments: notation and terminology}
 
  A counting experiment is one where we draw a random sample of $N$ items from some source and count the number of occurrences  of complementary events $A$ and $\neg A$ in that sample.  The count of occurrences, $k$, of the event of interest ($A$ or $\neg A$ as the case may be) is the result or `test statistic' in experiments of this type.  For the purposes of analysis we assume that $p_A$, the probability of a given item sampled from this source being an instance of $A$, is an independent and identically distributed variable ($p_A$ follows the same distribution for all items).  We also assume that we have no information about the occurrence of $A$ other than that obtained via sampling: more specifically, we assume that, with no sampling information, we must necessarily assume that every possible outcome $k$ is equally likely.   

To give a concrete example, consider a simplified version of Fisher's `tea tasting' experiment \citep{fisher1960design}. 
Suppose we have a colleague who claims to be able to tell, on tasting a cup of tea, whether or not the milk was poured into the cup before the tea.  To test this claim we set up a random sample of $N$ cups where the order of pouring for a given cup is determined by the toss of a coin, and recorded for each cup.  We control for any confounding factors that could potentially give clues about the pouring order by making sure that our colleague cannot see us pouring these cups, and by stirring each cup thoroughly, making sure all cups are the same temperature, making sure there are no any splashes of tea or milk around the cups, and so on.    We then ask our discriminating colleague to taste each of the $N$ cups and for each say whether they think milk or tea was poured first.  We record the number of incorrect responses $k$ out of $N$: this number is our test statistic. 

Note that the standard analysis of experiments of this type is based on the assumption that $p_A$ (the probability of successful identification, in our tea-tasting experiment) follows the same \textit{Bernoulli} distribution for all experiments (that $p_A$ has some fixed value $p$ for all sets of cups of tea).  Since counts of Bernoulli trials have a Binomial distribution, counting experiments analysed under this assumption are referred to as Binomial experiments.  We don't make this assumption here: we simply assume the probability $p_A$ follows \textit{some} distribution, and is independent and identically distributed across items.   

In the context of a counting experiment, we take an hypothesis $H$ to be a mechanism  which assigns probabilities to experimental results $k$, given sample size $N$.  We write these assigned probabilities as $p(k | N,H)$ (probability of result $k$ under hypothesis $H$ given sample size $N$).   
We take a null hypothesis to be any hypothesis under which the complementary events $A$ and $\neg A$ are  equally likely and so any deviation from equality in a sample must be a consequence of chance processes.  
A null hypothesis is thus necessarily symmetrical, assigning probabilities $p(k | N,H) = p(N-k | N,H)$ for all $k$.

It is useful to distinguish between one-sided and two-sided tests of a given hypothesis $H$.  In a one-sided test we are interested in whether experimental results deviate from those expected under chance in one specific direction.  Given a null hypothesis $H$, a particular significance level $\alpha$, and a particular sample size $N$, we take the critical value $K_{crit}$ of our statistic in a one-sided test to be the greatest value such that $p(k \leq K_{crit} | N,H) \leq \alpha$.   A result $k$ is then taken to be statistically significant (at significance level $\alpha$ in a one-sided test) when  $k \leq K_{crit}$. In our example tea-tasting experiment, for example, we are interested in whether the number of incorrect responses $k$ is lower than that expected by chance (whether our colleague is able to accurately identify the pouring order of milk and tea): a one-sided test.   If $k$   is lower than the critical value $K_{crit}$, then we conclude that our colleague's accuracy is greater than expected under our null hypothesis $H$.  

Given the symmetry of null hypotheses, it is clear that $p(k \leq K_{crit} | N,H) =p(k \geq N-K_{crit} | N,H)$ always holds.  This means that in a two-sided test (where we are interested in whether experimental results deviate  from those expected under chance, but don't care about the direction of deviation) we take the critical value $K_{crit}$ to be the greatest value such that $p(k \leq K_{crit} | N,H) \leq \alpha/2$.    A result $k$ is then taken to be statistically significant (at significance level $\alpha$ in a two-sided test) when  $k \leq K_{crit}$ or $k \geq N-K_{crit}$.  
 For simplicity of presentation we typically limit our discussion to the one-sided significance test   $k \leq K_{crit}$.

Finally  we define the probability of replication, $p_{rep}$, in terms of the test statistic values for two runs of a given experiment, with the same sample size $N$ used in each run and the same critical value $K_{crit}$.  Let $k_1$ represent the test statistic value for the first experiment and $k_2$ the value for the second; then $p_{rep}$ is equal to the probability of getting  $k_2 \leq K_{crit}$ given that $k_1   \leq K_{crit}$.  We write this as $ p_{rep}= p(k_2 \leq K_{crit}|N,k_1)$.   Since this measure assumes the same sample size in both experiments, it represents what is usually described in the literature as the `probability of exact replication' (the probability of getting a significant result in an exact replication of the original experiment).  Note that the question of replication of a statistically significant result necessarily involves a one-sided test for replication: a result $k_2$ replicates a result $k_1$ only if both are statistically significant and both deviate from chance in the same direction.

\section{Counting experiments and the Beta distribution}
\label{sec:countingExperiment}

A Binomial experiment is a counting experiment where we assume that each sampled item represents an independent and identically distributed Bernoulli trial, with a constant, fixed probability  $p_A$ of returning event $A$.  Under this assumption of a Bernoulli distribution for $p$, the probability of getting $k$ less than some value $K$ is given by the cumulative Binomial distribution 
\begin{equation*}
P(k \leq K  | N,p_A) =  \sum_{k=0}^{K} {N \choose k}p_A^k(1-p_A)^{N-k} 
\end{equation*}
We can generalise from this case by assuming that probability $p_A$ is not fixed, but varies from experiment to experiment according to a Beta distribution $p_A \sim beta(a,b)$.  The Beta distribution is, roughly speaking, the reverse of the Binomial distribution: where the Binomial distribution gives the chance of getting some number of successes $a=K$ (and failures $b=N-K$) given a fixed generating probability $p_A$, the Beta distribution gives the chance of the generating probability $p_A$ falling in a given range $p \ldots p+\Delta p$, given observed number of successes $a=K$ and failures $b=N-K$. 
 More specifically, if a generating probability $p_A$ follows a beta distribution $beta(a,b)$, this means that the chance of $p_A$ falling between $p$ and $p+\Delta p$ is given by
 \begin{eqnarray}
 \label{eq:beta}
 P( p \leq p_A \leq p +\Delta p;a,b)&=&\frac{p^{a-1}(1-p)^{b-1}}{
 	B(a,b)}\,\Delta p
 \end{eqnarray}
  where the normalising value  $B(a,b)$ is the Beta \textit{function},  defined for complex numbers $a$ and $b$ (whose real parts are positive) by
 \begin{eqnarray*}
 	B(a,b)&=&\int_0^1t^{a-1}(1-t)^{b-1}\mathrm{d}t
 \end{eqnarray*}
 which can be written in terms of the Gamma function as
  \begin{eqnarray*}
  	B(a,b)&= \frac{\Gamma(a)\Gamma(b)}{\Gamma(a+b)}
  \end{eqnarray*}
 giving the recurrence
\begin{eqnarray}
\label{eq:beta_recurrence}
  	B(a+1,b+1)&=& \frac{ab}{(a+b+1)(a+b)}B(a,b)
\end{eqnarray}
 
 Under the assumption  that the generating probability follows a Beta distribution $p_A \sim beta(a,b)$, the probability of getting $k$ less than some value $K$ is given by the cumulative Beta-Binomial distribution 
\begin{eqnarray}
\label{eq:beta_cumulative}
P(k \leq K |N, p \sim beta(a,b) ))	&=&\sum_{k=0}^{K}{N \choose k}\frac{B(k+a,N-k+b)}{B(a,b)}
\end{eqnarray}  
 A central property of the Beta distribution is that,  given that we have the prior distribution  $p_A \sim \mathrm{Beta}(a,b)$ and have also observed $k$ occurrences of $A$ in a sample of $N$ events, then the updated or posterior distribution for $p_A$ will be the Beta distribution 
\begin{equation} \label{eq:conjugate}
p_A \sim \mathrm{Beta}(a+k,b+N-k)
\end{equation} 
(a Beta prior necessarily gives a Beta posterior; the prior and the posterior are `conjugate', to use Bayesian terminology).     Note also that when $a=b$ the beta distribution is symmetrical (and so the cumulative  Beta-Binomial distribution is also symmetrical).  

It is useful to identify two limiting cases for the Beta distribution.  First, if $a = b = 1$ then 
\begin{eqnarray*}
	P( p \leq z \leq p +\Delta p)  &=&\frac{p^0(1-p)^0}{B(1,1)}\,\Delta p =  \Delta p
\end{eqnarray*}
and every possible value of $z$ is equally likely (the chance of $z$ falling in a given $p \ldots p +\Delta p$ range is simply equal to the size of that range): $beta(1,1)$ thus represents the uniform distribution.  Under the uniform distribution we have 
\begin{eqnarray*} 
	P(k |N, p \sim beta(1,1) ))	&=& \frac{1}{N+1}
\end{eqnarray*}
and all results $k$ are equally likely.   Second, for large values of the `shape' parameters $a$ and $b$, the Beta-Binomial distribution approximates the standard Binomial distribution with parameter $p = a/(a+b)$ arbitrarily well.  This means that the Beta-Binomial `contains' the Binomial distribution as a special case: a given Binomial distribution with parameter $p$ can be approximated to arbitrary precision by a Beta-Binomial distribution with $b = a(1-p)/p$, with the approximation approaching equality as $a$ tends to infinity.

Why do we use the Beta distribution, specifically, to represent generating probability distributions in counting experiments?  Recall that one of our assumptions for the analysis of counting experiments was that prior to sampling we we have no information about the occurrence of $A$, and so each outcome $k$ is assumed to be equally likely.  Since there are $N+1$ possible outcomes in an experiment with sample size $N$, these outcomes must have the same probability $1/(N+1)$, and we see that this assumption is equivalent to the assumption that, prior to sampling, $p_A \sim beta(1,1)$.  In other words, any counting experiment where information about the distribution of $A$ comes only from sampling, and so where all outcomes are assumed to be equally likely prior to sampling,  is a counting experiment where we can assume that prior to sampling  $p_A \sim beta(1,1)$.  Further, recall that another assumption for the analysis of counting experiments was that all information about the distribution of $p_A$ comes from sample counts.  From the `conjugate' property of the beta distribution (if the prior is a beta distribution, then all subsequent posterior distributions are also beta distributions), this in turn implies that all distributions for such counting experiments are beta distributions with integer parameters $a,b \geq 1$.

\section{Null hypotheses in  counting experiments}

A reasonable null hypothesis for any experiment is one which has some possibility of being true for that experiment.  We've seen that the only possible form for the unknown distribution of $p_A$ in a counting experiment (which we define as an experiment where we assume $p_A$ is an i.i.d variable, where information about $p_A$ comes only from sampling, and where without any sample information all values of $p_A$ must be considered equally likely)  is the Beta distribution; this means that all reasonable null hypotheses for experiments of this type must assign a Beta distribution for $p_A$.  We have also seen that the `initial' distribution, for all experiments of this type, is necessarily the uniform distribution $p_A \sim beta(1,1)$; this in turn means that all possible null hypotheses for experiments of this type must assign a distribution $p_A \sim beta(a,b)$ where $a \geq 1$ and $b \geq 1$.  

Recall that  in the context of a counting experiment, a null hypothesis is an hypothesis $H$ under which the symmetry $p(k | N,H)=p(N-k | N,H )$ holds for all $k$ and $N$.  The symmetry requirement only holds for beta distributions when those distributions themselves are symmetrical (when $a=b$), and so the set of all possible null hypotheses for counting experiments is equal to the set of all symmetrical Beta distributions $p \sim beta(a,a)$ for $a \geq 1$.   All such symmetric Beta distributions have the same mean, of $0.5$; the variance of such distributions around this mean falls as $a$ rises.   We  calculate the statistical significance of a given result $k$, relative to a given distributional null hypothesis $beta(a,a)$, as 
\begin{eqnarray}
\label{eq:beta_cumulative_null}
P(k \leq K |N, p \sim beta(a,a) )	&=&\sum_{k=0}^{K}{N \choose k}\frac{B(k+a,N-k+a)}{B(a,a)}
\end{eqnarray} 
and if this $P(k \leq K |N, p \sim beta(a,a) ) \leq \alpha$ then we say that our result is statistically significant (relative to that null).

What does it mean to take a beta distribution $p_A \sim beta(a,a)$ as a null hypothesis for a counting experiment?   The Binomial analysis of a counting experiment takes, as a null hypothesis, the proposal that the generating probability $p_A$ has a fixed point-form value of $0.5$ that does not change across experiments.  Taking a symmetrical beta distribution $p_A \sim beta(a,a)$ as a null hypothesis corresponds to proposing that the generating probability $p_A$ varies randomly and symmetrically around $0.5$, across experiments.  In our tea-tasting example, the Binomial analysis suggests as a null hypothesis the idea that our colleague can correctly identify pouring order at a rate of exactly $0.5$, across all cups of tea.  The beta distribution analysis, by contrast,  suggests as a null hypothesis the idea that our colleague's ability to correctly identify pouring order varies randomly around $0.5$ across different tea-tasting experiments, slightly higher in some experiments, and slightly lower in others (more specifically: varying randomly with the symmetrical  distribution of $beta(a,a)$ around $0.5$ ).

The proposal, that there is some degree of random variation in generating probability across experiments, reflects a fundamental and unavoidable aspect of the experimental process.   One central task for an experimenter is the control of confounding factors (factors which are of no theoretical interest, but which may influence the outcome of an experiment).  In our tea-tasting experiment confounding factors are  whether our  stirring distributed milk and tea evenly (pouring order could perhaps be deduced from poorly-stirred cups), whether the cups were of equal temperature (cups with tea poured in first will have a longer time to cool than those with milk first), whether our colleague could see us pour the cups, and so on. Our general null hypothesis is that responses will be random \textit{when these confounding factors are controlled}.   In fact, however, these confounding factors can never be completely controlled: the degree to which these confounding factors affect experimental results will vary randomly across experiments.  Variations in these confounding factors across experiments, in turn, causes variation in generating probabilities under the general null hypothesis.  In our tea-tasting example, we can never be $100\%$ sure that we have perfectly controlled all confounding factors (stirring, temperature etc) so that none of those factors give any indication of pouring order; the best we can say is that the association between these factors and correct identification of pouring order will vary randomly from experiment to experiment (to some greater or lesser degree) and so the probability of correct identification  under the null hypothesis will also vary randomly.    A null hypothesis which includes or allows for this random variation in confounding factors will be one which allows the underlying generating probability to vary from experiment to experiment.  Only a distributional-form null hypothesis can account for this form of variation; a  null hypothesis which assigns a fixed point-form value to the generating probability represents an unrealistic and overconfident assumption that all confounding factors are completely controlled and that the generating probability is constant across all experiments.    

We can consider the meaning of the `shape' parameter $a$ in these symmetric beta distributions in two ways.  First, the conjugate property means that beta distribution $beta(a+1,a+1)$ describes the normatively correct distribution of probabilities for some event $A$ given a sample of $2a$ items, half of which were $A$ (and where we have no other information about $A$).  The smaller this notional sample size $2a$ the more broadly these probabilities are distributed: the greater this sample size the more these probabilities are focused around $0.5$.  We can thus think of a beta distribution $beta(a+1,a+1)$ as representing the belief about $p_A$'s distribution we would have if we had seen a sample of $2a$ items, half of which were $a$ (and had no further knowledge about $A$).

Another way of interpreting this parameter $a$ is in terms of variance.  The variance of a Beta distribution $beta(a,a)$ is
\begin{equation}
\label{eq:beta_var}
\sigma^2 = \frac{1}{4(2a+1)}
\end{equation}   
and we see that low values of $a$ indicate a larger degree of variance around the mean (which is $0.5$ for all symmetrical distributions $beta(a,a)$), while high values of $a$ indicate smaller degrees of variance around the mean.  We can thus think of $a$ as an indicator of the `accuracy' of the distribution; the higher $a$ is, the more individual values of $p_A$ in samples will be tightly clustered around $0.5$.   It is worth noting that the variance of the  Beta-Binomial proportion distribution (the probability of $A$ occurring in a proportion $k/N$ of items in a sample of size $N$, given that $p_A \sim beta(a,a)$) is 
\begin{equation}
\label{eq:beta_Binomial_var}
\sigma^2 = \frac{1}{4(2a+1)} + \frac{a}{2N(2a+1)}
\end{equation}   
and so the Beta-Binomial proportion distribution that corresponds to a given Beta distribution $beta(a,a)$ always has a higher variance than that beta distribution, with the difference falling as sample size $N$ increases.

\section{Properties of distributional null hypotheses} 

At this stage it is worth noting two important properties of distributional null hypotheses for counting experiments.  First: there is an infinite set of possible null hypotheses (each corresponding to a different value of the shape parameter $a$) and hypotheses in this set are  nested within each other according to the value of $a$.   Suppose we have two null hypotheses $H_1: p \sim beta(a_1,a_1)$ and $H_2: p \sim beta(a_2,a_2)$ where $1 \leq a_1 \leq a_2$.  Both hypotheses describe symmetrical distributions  peaked  at $0.5$, with $H_2$ being more strongly peaked, less spread out, than $H_1$.   
Each null hypothesis gives a symmetrical unimodal probability distribution for results $k$, of $P(K |N, p \sim beta(a_1,a_1) )$ for $H_1$ and $P(K |N, p \sim beta(a_2,a_2) )$  for $H_2$, with the distribution associated with $H_2$ being more strongly peaked (just as $H_2$ itself is more strongly peaked).  This means that if some value $K$ has a cumulative probability less than $\alpha$ for $H_1$ (is a significant result at level $\alpha$ relative to that null hypothesis), then that value $K$ will necessarily also have  a cumulative probability less than $\alpha$ for $H_2$ (be a significant result at level $\alpha$ relative to that null hypothesis).  In other words, if a given result is statistically significant relative to a given null hypothesis $H: p \sim beta(a_1,a_1)$, that result is also statistically significant relative to all null hypotheses $H: p \sim beta(a_2,a_2)$ when $a_2 > a_1$.  When reporting the statistical significance of an experimental result at level $\alpha$, then, it makes sense to report this relative to the some low value of the parameter $a$ for which
\begin{eqnarray} 
\label{eq:K_crit}
	P(k \leq K |N, p \sim beta(a,a) )	\leq \alpha
\end{eqnarray} 
since significance relative to the distributional null $ beta(a,a) $ at level $\alpha$ implies significance  at the same level relative to all other nulls $beta(a',a')$, where $a' > a$.  Further, since a  Binomial distribution with parameter $p$ is approximated to arbitrary precision by a Beta-Binomial distribution with $b = a(1-p)/p$ (with the approximation approaching equality as $a$ tends to infinity), we see that the Binomial distribution with $p=0.5$ is exactly equal to the Beta-Binomial distribution $P(K |N, p \sim beta(a,a) )$ as $a \rightarrow \infty$.  This means that if a given result is statistically significant relative to a some distributional null hypothesis $H: p \sim beta(a,a)$, it is also statistically significant relative to the Binomial distribution with $p=0.5$.  

This first property concerns the relationship between symmetrical beta distributions, and beta-Binomial distributions, for different values of the shape parameter $a$.  The second property relates the beta and beta-Binomial distributions for the same value of the shape parameter.  It turns out that if the generating probability $p_A$ is distributed across samples according to some beta distribution $ beta(a,a) $,then as sample size $N$ goes to infinity, the distribution of sample proportions $k/N$ approaches that of the underlying probability distribution $ beta(a,a) $. More precisely, defining the regularised incomplete Beta function
$I(r; a,b)$ to represent the normalised area under the Beta distribution from $0$ to $r$, we have
\begin{eqnarray} 
\label{eq:bound}
	P(k \leq K |N, p \sim beta(a,a) )	\geq I(K/N; a,a)  + \epsilon
\end{eqnarray}
for all $K/N \leq 0.5$, with the difference $\epsilon$ always being positive and declining to $0$ as $N$ approaches infinity.  To see informally why this relationship holds, simply consider that as $N$ goes to infinity the variance of the sample proportion (the Beta-Binomial proportion variance, Equation \ref{eq:beta_Binomial_var}) declines towards the variance of the population proportion (the Beta variance, Equation \ref{eq:beta_var}).  The Appendix gives a more formal proof of this result.

    This result means that  the critical value $K_{crit}$  for a distributional null hypothesis at a given sample size $N$, as defined in Equation \ref{eq:K_crit}, is always less than some proportion $zN$ of the sample size.  To illustrate this point, suppose that, for a given significance level $\alpha$ and  distributional null $ beta(a,a) $,  we find the value $z$  such that $I(z; a,a) = \alpha$.   From Equation \ref{eq:bound} we see that $P(k \leq K_{crit} |N, p \sim beta(a,a) ) \leq \alpha$ can only hold if $I(K_{crit}/N; a,a)  + \epsilon \leq \alpha$.   But since $I(z; a,a) = \alpha$, this can only hold if  $K_{crit}/N \leq z$; or equivalently, if $K_{crit} \leq zN$.  The critical value $K_{crit}$ is thus always less than a constant proportion $z$ of the sample size $N$, and approaches $zN$ as $N$ goes to infinity.  For a one-sided test, this result means the rejection region for the null hypothesis  $ beta(a,a) $ is always smaller than $zN$: a result $k \geq zN$ will never become statistically significant, no matter how large the sample size $N$.  For a two-sided test with distributional null $ beta(a,a) $ we find $z$ such that $I(z; a,a) = \alpha/2$; then  the rejection region for the null hypothesis is always smaller than the region $[0,z] \cup [1-z,1]$ and results $k$ where $zN < k < (1-z)N$ will never be statistically significant, irrespective of the sample size $N$.

\section{Addressing problems with point-null hypothesis testing}

This distributional approach  addresses a range of problems with the standard point-form NHST approach.
The first problem with the point-form approach is a consequence of the relationship between probability distributions and sample proportions: as sample size $N$ goes to infinity, the distribution of sample proportions approaches the underlying probability distribution.  With a point form distribution, this means that  as sample size $N$ goes to infinity, the probability of getting a sample proportion not equal to the point-form probability value falls to zero (and so any sample proportion not exactly equal to the point-form value will have a very low probability of occurrence and so be counted as statistically significant).  In other word the probability of getting a statistically significant result under a point-form null hypothesis is simply an increasing function of the sample size $N$: the larger the sample size $N$ the more likely we are to get a statistically significant result, and we are guaranteed to get a significant result at \textit{some} sample size.

This aspect of point-form null hypotheses is troubling for a number of reasons.  Most importantly, it undermines the connection between the statistical significance of an experimental result and the importance of that result.  The fundamental aim, in NHST, is to distinguish between real effects and results that are due to random chance; this is done by taking statistical significance (low probability under the uninteresting null hypothesis) as an indication that the result is real.  Under point-form null hypotheses,  however, statistical significance can be achieved  in any experimental setting by simply taking a large enough sample size.  If any result can be made statistically significant by increasing sample size, then all results are `real'. More practically, this problem means that researchers can produce statistically significant results (relative to a point-form null hypothesis), by simply throwing resources at the experimental process (that is, by using very large sample sizes). Finally, this problem seems to potentially undermine fixed-effect meta-analytic results, often taken as as more reliable than individual studies: the problem being that a fixed-effect meta-analysis essentially corresponds to a much larger sample, and so has a greater chance of showing a statistically significant effect \textit{purely due to the increase in sample size}.  We can see this problem in action when we consider cases where the constituent experiments making up a meta-analysis are not themselves statistically significant to any great degree, but the meta-analysis overall produces a significant result \citep[see, e.g.][]{braver2014continuously}.  

This problem does not arise with distributional, as opposed to point-form, null hypotheses.  In this case the distribution of sample proportions again matches the null hypothesis distribution as $N$ goes to infinity, but because the distributional null hypothesis  assigns a non-zero value to all probabilities (and so all proportions), the probability of getting a given sample proportion will never fall to zero: instead, as we saw above, the probability of getting a sample proportion $k \leq zN$ approaches, in the limit, the probability of getting a generating probability $p_A \leq z$ under the distributional null hypothesis.   With a distributional-form null hypothesis, in other words, the probability of getting a significant result for a given sample proportion is not a function of the sample size $N$, but of the probability assigned to that proportion by the underlying null hypothesis.

This issue, that with a point-form null hypothesis the probability of achieving a significant result rises with sample size, underlies another commonly made objection to NHST: that the null hypothesis is always false.  To quote \citet{cohen2016earth}
\begin{quote}``[the null hypothesis ] can only be true in the bowels of a computer processor running a Monte Carlo study (and even then a stray electron may make it false). If it is false even to a tiny degree, it must be the case that a large enough sample will produce a significant result and lead to its rejection. So if the null hypothesis is always false what's the big deal about rejecting it?"
	\end{quote}
This objection, again, only applies to the point-form null hypothesis, and not to distributional-form nulls, and for the same reason: the probability of falsification of a distributional-form null is not a function of sample size $N$ but of the continuous probability distribution itself.  A distributional-form null is not `always false', and so rejection of such a null hypothesis is always meaningful.  As we saw above, if for distributional null hypothesis $H:beta(a,a)$ and significance level $\alpha$ we have $I(z;a,a)=\alpha/2$, then in a two-sided test the null hypothesis will not be rejected for results $zN<k <(1-z)N$, irrespective of sample size $N$.  In a two-sided test against a distributional null hypothesis  the `rejection area' increases to a fixed proportion $2z$ of the outcome space.  For a point-null hypothesis, by contrast, the rejection area in a two-sided test increases monotonically with $N$ to a maximum of $1$ (with large enough $N$, all results are in the rejection area for a point-form null hypothesis).

In response to this relationship between statistical significance and sample size for point-form null hypotheses, a number of researchers have stressed the use of effect size $d$ as well as statistical significance $p$ in reporting results.  The idea here is that we may have phenomena that are `real' (producing results that are statistically significant relative to the point-form null), but that some of those phenomena have negligible observable effects (very small differences $d$ between the observed result and the point-null value).  By reporting both $p$-value and effect size $d$, readers are made aware of this, and can presumably focus their attention on phenomena that are `really real': not just statistically significant relative to the point null, but also having real observable effects.  The problem with this approach is that it leaves us with no guidance as to what counts as a `real observable effect': it does not tell us where we should draw the boundary between negligible and `real' effect sizes.  Again, this problem does not arise with distributional null hypotheses.  With a distributional null hypothesis there is a direct relationship between statistical significance and effect size: only effect sizes that are larger than a certain value will ever reach statistical significance, independent of sample size.  To see this, consider that in a counting experiment a natural measure of effect size, for a given result $k$ and sample size $N$, is the difference between the proportion $k/N$ and the expected value under the null hypothesis, $0.5$.  If for distributional null hypothesis $H:beta(a,a)$ and significance level $\alpha$ we have $I(z;a,a)=\alpha/2$, then a result $k/N$ will be statistically significant in a two-sided test only when $k/N \leq z$ or $k/N \geq 1-z$: only effect sizes of $z$ or greater will ever reach statistical significance.

A further problem is connected to the use of point-form NHST in confirmatory research.  A core principle in confirmatory research is the need to be conservative when making decisions about rejecting the null hypothesis (the hypothesis that experimental results are due to chance).  Point-form null hypotheses, however, are not conservative about the effects of chance on experimental results.  We can see this in two ways.  First, compared to distributional-form null hypotheses, point-form nulls will only accept a narrow range of sample proportions as consistent with chance (a range that shrinks as sample size $N$ rises: see our first point above).  Distributional-form nulls accept a wider range of sample proportions as consistent with the null (a range that does not shrink as $N$ rises).  This means results that would support rejection of the null under a point-form null hypothesis will not support rejection under a distributional-form null: the use of distributional-form null hypotheses makes the researcher more conservative about rejection of the null.  The same point follows from our discussion of variability in control of confounding variables: point-form null hypotheses are overconfident in that they assume perfect control of these variables in all experiments, while distributional-form nulls are more conservative in that they allow for this control to vary randomly across experiments.  

All of the above problems arise because of the point-form null assumption that the generating probability underlying performance doesn't vary from experiment to experiment, but is fixed.  This is clearly not a realistic model of the experimental process, where unavoidable confounding factors have effects that vary randomly across experiments and so cause variation in the generating probability across those experiments.  A distributional-form NHST approach is necessary to address the unavoidable aspect of experimental design.   

 Our final problem concerns the `replication crisis': the finding that a large proportion of statistically significant results cannot be replicated in subsequent studies.    In response to this crisis, scores of articles have been written that diagnose the problem, evaluate its impact, and suggest solutions.  Numerous explanations for this failure to replicate have been given, including questionable research practices such as `p-hacking', `HARKing' (hypothesizing after results are known) `file-drawer' effects, deception around experimental design, and explicit fraud.    The above discussion of distributional nulls gives a straightforward statistical account for these replication results: these replication failures arise because standard NHST analyses in terms of a point-form null are systematically overconfident in terms of rejecting the null (leaving out a range of factors that may cause random experiment-to-experiment variation).  This means that many results that are counted as statistically significant (against a point-form null) are, in fact, not real effects  and so will not replicate.  

This distributional-null account of replication goes beyond accounts that see failure to replicate as a reflection of questionable research practices.  In particular, this distributional-null account makes a general prediction about the extent to which replication will vary across fields of research.  One problem with point-form nulls is that they do not take into account random variation in control of extraneous or confounding factors across experiments.  In fields where such extraneous confounding factors are rare, or are easily controlled, the point-form null will be relatively close to the distributional null that accounts for variation in confounding factors.  In such fields we would expect most results that are significant against a point-form null to replicate.  In fields where confounding factors are frequent and harder to control, however, the point-form null will be substantially different from the distributional null that accounts for variation in those factors.  In fields with frequent and hard to control confounding factors, then, we would expect most results that are significant against a point-form null to fail to replicate.   Data on replication rates across fields support this proposal to some extent, with replication being a problem in areas such as psychology, neuroscience and medicine (areas focused on very complex and only partly understood systems, and so necessarily affected by many confounding factors). Within-field comparisons support this proposal with, for example, higher replication rates in Cognitive Psychology experiments (where variation in cognitive factors is the primary confound) than Social Psychology experiments\citep[where variation in cognitive, interpersonal and societal factors all act as confounds; see e.g.][]{open2015estimating}.  

\section{Estimating the probability of replication}

The previous section showed how the use of distributional null hypotheses naturally resolves various problems with standard point-form NHST.  In this section we show that the use of distributional nulls adds to our statistical toolbox, allowing us to directly and coherently estimate the probability of replication of experimental results.  

Estimation of this probability of replication (which write as $p_{rep}$) is problematic in the standard point-form NHST approach. The standard approach is via the concept of statistical power \citep[e.g.][]{greenwald1996effect,posavac2002using,gorroochurn2007non}. This approach involves two contrasting point hypotheses: the point null hypothesis $H_0$, and an alternative hypothesis $H_A$, whose point value corresponds to the result observed in experiment $1$.    This approach assumes that, since a significant result has been obtained in experiment $1$,  we should conclude that the null hypothesis $H_0$ is probably false and the alternative hypothesis $H_A$ is probably true (in other words, this approach assumes that the population effect size is equal to the result, or effect size, seen in experiment $1$).  Replication is then measured in terms of the probability of getting a statistically significant result under the assumption that $H_A$ is true\footnote{An alternative approach, given by \citet{killeen2005alternative},  estimates the probability of getting a result in experiment $2$ that goes in the same direction as that seen in experiment $1$, whether statistically significant or not.   Since this approach doesn't address replication of statistically significance, we don't consider it here.}.

In terms of a counting experiment, these two point hypotheses would be  $H_0: P(A)=0.5$ (the `point null' hypothesis) and $H_A: P(A)=k/N$ (the alternative point hypothesis, that the population probability of $A$ is equal to the proportion $k/N$ observed in experiment $1$).  The critical value $K_{crit}$ is obtained relative to the null hypothesis $H_0$, and the probability of replication is the probability of getting a value less than $K_{crit}$ under the hypothesis $H_A: P(A)=k/N$; a value given by the cumulative Binomial sum
\begin{equation*} 
p_{rep} = p(k_2 \leq K_{crit}| N,  H: P(A)=k/N) = \sum_{k_2=0}^{K_{crit}} {N \choose k_2} \left(k/N\right)^{k_2} \left(1-k/N\right)^{(N-k_2)}  
\end{equation*}

This approach suffers from  a fundamental problem: it is based  on the difficult-to-justify assumption that the expected value of the test statistic  in a replication will vary around a value equal to the observed value of the test statistic in the first experiment.   \citet{greenwald1996effect} suggest this issue should be dealt with using a Bayesian approach which considers the prior probability distribution for the population parameter. \citet{macdonald2005replication} makes the same point.   To see why such a prior distribution is required, consider that to estimate $p_{rep}$ we must have an expression for the variability of the test statistic across replications of a given experiment.  Variability of a result across replications cannot be estimated from a single value (from a single experimental result), and so some form of  distribution for the population parameter in question is necessary,  if we are to give a realistic estimate for the probability of replication.    

 The distributional-null approach gives us a natural way of estimating the probability of replication of a statistically significant result in a counting experiment, relative to some distributional null hypothesis $beta(a,a)$.  Suppose we have carried out our counting experiment, and have seen $k_1$ occurrences of $A$ in a sample of size $N$.        Suppose this first experimental result was statistically significant: $k_1 \leq K_{crit}$ for some critical value $K_{crit}$ .  If we now repeat this experiment with the sample size $N$, what is the probability of a result at the same significance level: a second result that is also less than  $K_{crit}$, given the same sample size?  Since observation of this result $k_1$ leads to an updated distribution for $p_A$ of $p \sim beta(k_1+a, N-k_1+a )$  this probability is given by

\begin{eqnarray}
\label{eq:replication}
p_{rep} &=&  
P(k \leq K_{crit} |N, p_A \sim beta(k_1+a, N-k_1+a ))  \nonumber \\
	  &=&\sum_{k=0 }^{K_{crit}}{N \choose k}\frac{B(k_1+k+a, N-k_1+N-k+a)}{B(k_1+a, N-k_1+a)}
\end{eqnarray}  
with the second line following from Equation \ref{eq:beta_cumulative}.  This expression gives the probability of replicating a statistically significant result in a counting experiment, given the observed result $k_1$ in the first experiment (and given the distributional null hypothesis $beta(a,a)$).  

It is useful to consider this model of replication probability in the specific situation when $k_1= K_{crit}$.  For the statistical power models described earlier, $p_{rep}$ is necessarily $0.5$ in this situation (because those models assume that the result $k_1$ in the first experiment represents the true population value of that statistic, and so replication values will vary randomly around that value).  The same result holds in this model, but only with the uniform null hypothesis, and for essentially the same reason: using the uniform null hypothesis ($a=1$) corresponds to the assumption that the result $k_1$ in the first experiment represents the true population value of that statistic.    For other distributional null hypotheses (where $a>1$) we necessarily have $p_{rep} < 0.5$ in the specific case where $k_1 = K_{crit}$, with $p_{rep}$ falling as $a$ rises (because as $a$ rises the null distribution comes to dominate the observed result $k_1$).   

These results means that if the $p$-value in our original result was close to the boundary of significance  $\alpha$, then the probability of replicating that result (the probability of getting a $p$-value  less than $\alpha$ in a replication) is necessarily not going to be much above $50\%$, and in fact will mostly be below that bound. This result is typically supported in replication research: results with significance close to $\alpha=0.05$ will typically replicate, at the same significance level $\alpha=0.05$, at a rate of somewhere around $30\%$ or $40\%$.   This is is some ways bad news: it means that many statistically significant results are unlikely to replicate at any significant rate.  A more helpful perspective, however, is to see this as simply a mathematical consequence of random variation in repeated experiments and to adjust our expectations of replication accordingly, accepting the mathematical fact that results with significance levels close to $\alpha$ are  unlikely to produce statistically significant replications at the same level.

  Ideally we would like to be able to say that a given experimental result is `real' both in terms of  statistical significance (a result is real because it is unlikely to be due to chance processes) \textit{and} in terms of replication (a result is real because it is experimentally demonstrable in repeated experiments).   To express this concretely: we would like to count an experimental result as real when that result is statistically significant at criterion $\alpha$ (relative to our null hypothesis $beta(a,a)$) and where its probability of replication is greater than some criterion $\beta$.  Equations \ref{eq:beta_cumulative} and \ref{eq:replication} allow us to calculate these probabilities.  In the next section we discuss the practical use of these calculations.
  
 \section{ Using distributional null hypotheses in NHST}
 
The above has hopefully convinced readers that a distributional NHST approach has much to recommend it.  At this point we outline one way of using this approach in practice.  As above, we assume a counting experiment, and consider symmetrical beta distributions with $a \geq 1$ as possible distributional-null hypotheses. 
 
We are interested in asking whether experimental results give evidence for the existence of a real effect.  `Real' has two meanings here: an effect is `real' if the result demonstrating that effect is unlikely to be due to chance (has a low probability under the null hypothesis) and also, a effect is `real'  if the result demonstrating that effect is likely to be repeated in future experiments (has a high probability of being observed in a replication).  We can set subjective criteria for both these measures, and say that we will take a result as giving evidence of a real effect if its $p$-value (under a given null hypothesis) is less than some value $\alpha$ AND if its probability of replication $p_{rep}$ under that null hypothesis is greater than some level $\beta$. More formally, for a given null hypothesis $beta(a,a)$, sample size $N$, and subjective criteria $\alpha$ and $\beta$, the statistical significance of a given result $k_1$ is
   \begin{equation*}
 p_{sig} = P(k \leq k_1| N, p \sim beta(a,a)) 
   \end{equation*} 
Identifying the critical value, $K_{a}$, associated with null hypothesis $beta(a,a)$ as the maximum value  such that 
 \begin{equation*}
 P(k \leq K_{a}| N, p \sim beta(a,a)) \leq \alpha
 \end{equation*}
 we see that the probability of a significant result in a replication given this null hypothesis  is 
    \begin{equation*}
    p_{rep} = P(k \leq K_{a} |N, p \sim beta(k_1+a, N-k_1+a )) 
    \end{equation*} 
Combining these measures an experimental result $k_1$ can be counted as a real effect, relative to our criteria  $\alpha$ and $\beta$, in a one-sided test when
\begin{eqnarray}
\label{eq:real_effect}
p_{sig} \leq \alpha  \textit{ \ AND \ }   p_{rep}   \geq \beta
\end{eqnarray}
 holds for some null hypotheses  $beta(a,a)$.  If Equation \ref{eq:real_effect} does not hold for any null hypotheses $beta(a,a)$, then our result $k_1$ cannot be counted as a real effect relative to our subjective criteria.   If Equation \ref{eq:real_effect} does hold for some $beta(a,a)$ we can say that our result $k_1$ could reflect a real effect.

We can apply this approach generally as follows.   Assuming a one-sided test with  criteria $\alpha$ and $\beta$, and sample size $N$  we calculate a threshold $K_{bound}$ equal to the highest value $k_1$ such that Equation \ref{eq:real_effect} holds for some null $beta(a,a)$, if such a threshold exists.  If no such threshold exists, then no results from our experiment can ever reflect a real effect relative to our criteria. If such a threshold does exist, than any results $k > K_{bound}$ are inconsistent with a real effect relative to our criteria;  results $k \leq K_{bound}$ are consistent with a real effect, to a degree that depends on the value of $k$.  

To produce a tractable computational algorithm for calculating this threshold $K_{bound}$  for given $\alpha$, $\beta$, and $N$, we note that critical values $K_a$ for null hypotheses $beta(a,a)$ can never exceed the critical value for Binomial null hypothesis, which we write as $K_{bin}$ and which is the maximum value such that
\begin{equation*}
 P(k \leq K_{bin})| N, p=0.5) \leq \alpha
\end{equation*}
We also note that for fixed $K_{crit}$  the relationship
    \begin{equation*}
P(k \leq K_{crit} |N, p \sim beta(k_1+a, N-k_1+a )) \geq  P(k \leq K_{crit} |N, p \sim beta(k_1+(a+1), N-k_1+(a+1) ))
    \end{equation*} 
must hold, because the $a \rightarrow a+1$ increase on the right hand side acts both to reduce the variance of the associated beta distribution and move that distributions centre towards $0.5$, with both changes reducing the the tail of the distribution (and so reducing the chance of getting a value less than $k_{crit}$).   This in turn means that, for fixed $K_{crit}$, the probability of replication $p_{rep}$ given distributional null $beta(a,a)$ necessarily falls as $a$ increases.  Together these two points mean that if we have some $a^*$ for which $p_{rep} \geq \beta$ does not hold and where  $K_{a^*} = K_{bin}$, then we can conclude that  $p_{rep} \geq \beta$ will similarly not hold for any $a > a^*$, and so Equation \ref{eq:real_effect} will not hold for $a \geq a^*$. Our search for $K_{bound}$ (for values of $k$ which do satisfy Equation \ref{eq:real_effect}) is thus computationally tractable because it need only consider values of $a$ less than $a^*$.  Code implementing this search  is available in R and via a web interface at \url{http://inismor.ucd.ie/sigrep/}. 
For arbitrary values of $N$,$\alpha$ and $\beta$ this  code returns $K_{bound}$ (if it exists) or \textit{nil} if such a bound does not exist.
 
If we find that $k_1 >K_{bound}$, then our result cannot in any way be taken to indicate a real effect relative to significance and replication criteria $\alpha$ and $\beta$, while if $k_1 \leq K_{bound}$ the best we can say is that  $k_1$ may reflect a real effect, relative to these two criteria.  The situation here is exactly analogous to that in standard significance testing, where a result $k_1$ that is not statistically significant cannot be taken to indicate a real effect (relative to the significance criterion $\alpha$), while if $k_1$ is statistically significant, the best we can say is that this result may reflect a real effect.

  Consider, for example, the following results in a `tea-tasting' experiment. Suppose we have a significance criterion of $\alpha = 0.05$ and a replication criterion of $\beta = 0.5$.  We present our colleague with $8$ cups of tea, and they identify the pouring order correctly in all cases ($N=8,k=0$).  Under a point-form null hypothesis (that is, under the Binomial distribution with $p=0.5$) this result has a $p$-value of $p \sim 0.004$ and so is statistically significant according to our criterion.  Taking inputs $N=8$, $\alpha = 0.05$ and $\beta = 0.5$ and carrying out the search algorithm mentioned above, however, we find that no bound $K_{bound}$ exists: no result from this experiment can ever meet our criteria for significance and replication.
  s
  To illustrate this result in detail we consider distributional null hypotheses $p \sim beta(a,a)$, and find that this result $N=8,k=0$ is significant at $\alpha = 0.05$  for null hypotheses $beta(a \geq 3)$. Calculating the probability of a statistically significant replication given  $N=8,k=0$ and the prior assumption $a = 3$ (the minimum assumption we must make if we are to count this result as statistically significant in the first place) we get $p_{rep} = 0.21$ (only a $1$ in $5$ chance of replication).  Since $p_{rep}$ falls as $a$ rises, higher values of $a$ will give lower values of  $p_{rep}$ (for $a=4$ we get $p_{rep} = 0.15$, for $a=5$ we get $p_{rep} = 0.12$, and so on).   These values $(p \leq \alpha = 0.05, p_{rep} = 0.21, H_0=beta(a =3))$ thus tell us that there is no possible null hypothesis under which our result is both statistically significant (at level $\alpha = 0.05$) and replicable (at $p_{rep} \geq \beta = 0.5$).  
  
  Suppose we now present our colleague with $16$ cups of tea rather than $8$ (doubling the sample size), with the same criteria $\alpha = 0.05$, $\beta = 0.5$.  Calculating as above we get $K_{bound}=0$, and so results $k_1 = 0$ will satisfy our criteria for some $a$.     With $\alpha =0.05$ the result $k_1=0$ is significant relative to all null hypotheses $beta(a \geq 4)$; and taking $a=4$ we get $p_{rep} \sim 0.52$.   These values $(p \leq \alpha = 0.05, p_{rep} \sim  0.52, H_0=beta(a = 4))$ tell us that our result is  significant at $\alpha = 0.05$, and that there is a more than $50\%$ chance that a replication will also give a significant result at that level (given the null hypothesis $H_0=beta(a = 4)$).  Since the statistical significance of a given result increases as $a$ increases, we can thus say that our  $N=16,k=0$ result is statistically significant under almost all possible null hypotheses (all null hypotheses $beta(a \geq 4)$), and so is inconsistent with almost all possible null hypotheses.  Since probability of replication $p_{rep}$ falls as $a$ increases, we can say that the probability of replication of our result (assuming any of the remaining null hypotheses $beta(a < 4)$ that are consistent with that result ) is greater than $0.5$.  Our result $k=0$ thus may reflect a real effect relative to our significance and replication criteria $\alpha$ and $\beta$.

\section{Discussion}

Our argument in this paper is for a move away from the standard point-form approach to NHST  to  a broader distributional-form NHST approach.  Standard point-form NHST suffers from various fundamental problems: the probability of getting a statistically significant result in the standard NHST  increases with sample size, irrespective of the presence or absence of a true effect; the standard NHST approach is overconfident in identifying effects as real; the standard NHST approach does not allow for correct estimates of the probability of replication of a given result. A distributional-form approach to NHST avoids problems associated with  sample size,  allows for conservative rejection of the null, and allows meaningful and coherent estimation of the probability of replication. 

\subsection{Possible objections}

We expect a number of objections to our argument.  The first objection concerns the use of a  distributional representation of the null hypothesis. `These distributional nulls are just a Bayesian priors in another form' we imagine the objection goes, `and Bayesian priors are subjective measures of belief, not objective probability estimates.  Subjective beliefs cannot enter into objective frequentist hypothesis testing'.  

It is true that our distributional nulls have a mathematical form that is identical to a Bayesian prior.  It is not true, however, that these distributional nulls represent subjective measures of belief.  Instead, these distributional nulls play the same  role that point-form nulls play in NHST: for a given (distributional or point-form) null hypothesis $H$ we say `result $k$ would have a low probability of occurrence if null hypothesis $H$ were true, and so result $k$ is unlikely to be simply a consequence of random processes'.  Such  assertions do not require or reflect any subjective belief in the null hypothesis.  To put this response another way: since the logic of NHST is independent of the form of null hypothesis being used, our use of distributional rather than point nulls does not change the objective nature of such hypothesis testing.

A second objection concerns statistical testing against distributional null hypotheses characterised by a free parameter $a$ (representing the precision of the particular null distribution being used).  `Researchers can adjust this null hypothesis distributional parameter  $a$ until they find a null hypothesis against which their observed results are statistically significant' we imagine the objection goes. `But this is simply a form of data-dredging or `p-hacking': an attempt to find patterns in data that can be presented as statistically significant when in fact there is no real underlying effect'.  

Our response here is to note that the use of distributional null hypotheses systematically reduces the occurrence of statistically significant results, relative to the point-form null.  This is because, as we saw earlier, the rejection region for a null hypothesis grows monotonically with its precision (with $a$): and so the maximum rejection region (and so the greatest chance of a statistically significant result) arises with the point-form null.  Given this, a better characterisation of this process of testing against nulls characterised by  this parameter $a$ is one where we attempt to \textit{reduce} the chance of finding statistically significant results: where we are conservative in accepting patterns in data as being statistically significant.

A third possible objection concerns the availability of alternative hypothesis-testing methods, such as the Bayes Factor test (commonly put forward as a replacement for NHST).  `While the NHST approach was appropriate in the last century', we imagine the objection goes, 'today we have better statistical approaches to hypothesis testing; we don't need the NHST approach'. 

 We have two responses to this.   The first is to point out that the Bayes Factor approach, at least as it is typically used, is also based on the assumption of a point-form null hypothesis, and as such falls prey to many of the problems described above (problems to do with sample size, replication and so on).  The second, more general response is to point out that the NHST and Bayes Factor approaches ask two different questions.  The NHST approach involves a single null hypothesis  $H_0$ (the hypothesis that observed results are the chance consequence of some random process) and asks whether observed results are likely or unlikely under that null hypothesis.     The Bayes Factor approach, by contrast,  involves comparison of two contrasting hypotheses: Bayes Factor analysis asks whether experimental results give evidence for one hypothesis $H_0$ or for a specified alternative hypothesis $H_1$.   Both forms of question are useful and important: one, however, does not replace the other.
 
\subsection{Related approaches}

While the idea of using distributional rather than point-form null hypotheses will, we think, be relatively novel to most researchers in psychology and the cognitive science generally, it is worth pointing out that similar approaches are well known and commonly used in certain specific areas. One such area is the study of the `sex ratio' (or equivalently, the proportion of female children) in families.  Across the human population in general, the sex ratio is close to $1:1$ (the probability of a randomly selected child being female is almost exactly $0.5$).  The distribution of male and female children in families, however, does not follow the Binomial distribution with $p=0.5$.  Instead, this distribution is `overdispersed' relative to the Binomial, with some families having more female children, and some more male children, than would be expected from the Binomial model.      
A Beta-Binomial model, where $p(female)$ varies from family to family according to a beta distribution, gives a good fit to the family sex ratio distribution , and it is often assumed as a model of the sex ratio \citep{lindsey1998analysis}.

  Perhaps the most important of these is that of `random effect meta-analysis'.  Meta-analysis attempts to assess the reality and size of some effect $d$ by systematically combining measures of that effect $d_i$ from different experimental studies.     A ‘fixed effect’ meta-analysis starts with the hypothesis that the true value of this effect $d$ is fixed at some point-form value, and that individual measures $d_i$ represent samples from a population with that fixed point-form value.  A `random-effect’ meta-analysis, by contrast, starts with the hypothesis that the effect $d$ has a distributional, rather than a point, form (this distribution is the random effect) and that individual measures $d_i$ represent samples from that distribution.   The idea that a distributional, rather than point-form, model of effects should be used in the medical and social sciences is well understood in the meta-analysis literature: to quote \citep{higgins2009re}
\begin{quote}
Occasionally it may be reasonable to assume that a common
effect exists (e.g. for unflawed studies estimating the same physical constant). However, such an
assumption of homogeneity can seldom be made for studies in the biomedical and social sciences.
 These studies are likely to have numerous differences, including the populations that are
addressed, the exposures or interventions under investigation and the outcomes that are examined. Unless there is a genuine lack of effect underlying every study, to assume the existence of
a common parameter [a fixed effect] would seem to be untenable.  
\end{quote}
There are, however, a number of major differences between the `distributional effect $d$' model used in random-effect meta-analysis and the `distributional null hypothesis' approach we describe.  Most obviously: our approach tests experimental results against a distributional null hypothesis, and does not involve any assumptions about the `true effect' $d$ underlying those results.  Random-effect meta-analysis, by contrast, makes no mention of the (point-form or distributional) null, and instead involves distributional assumptions about the effect size $d$, based on experimental results $d_i$.  
In some ways this approach is similar to the statistical power model of replication described above, which begins with the assumption that the population effect size $d$ is equal to the observed result $d_1$ (disregarding the possibility that the observed result could have been produced under the null hypothesis).

\section{Conclusions}

 Standard point-form NHST has a number of fundamental mathematical, experimental and practical problems.  These problems can be effectively and naturally addressed by moving to a distributional NHST approach.  The primary difference between the distributional and point-form NHST approach lies in the fact that the distributional approach accounts for random variation across experiments, due to the unavoidable presence of randomly varying confounding factors.  Point-form NHST, because it does not account for this type of random distributional variance, is overconfident in rejecting the null: results which arise purely as a consequence of random distributional variance  will be counted as statistically significant under point-form NHST. 

 The extent of this overconfidence depends on the number of confounding factors affecting experimental results: experiments in areas such as psychology, neuroscience, medicine, and so on (areas investigating very complex, interacting, and only partially understood systems) will necessarily be subject to many potential confounds, and so point-form judgements of statistical significance in these areas will be substantially and systematically overconfident: many results that are counted as real relative to standard significance criteria ($\alpha = 0.05$ or $\alpha = 0.01$, for example) are in fact likely to have arisen as a consequence of random distributional variance.   
 
 A second problematic aspect of point-form NHST arises from the fact that the probability of getting a statistically significant result in the point-form NHST increases with sample size, irrespective of the presence or absence of a true effect.  This means that even in situations where confounding factors are very tightly controlled (and so distributional variance minimised) the point-form NHST approach will still be systematically overconfident, with results being judged statistically significant simply as a consequence of a large enough sample size, rather than a true effect.   This issue is particularly important in areas involving hypothesis testing across large data sets (again, medicine and to some extent psychology), where again results that are counted as real relative to standard significance criteria are in fact likely to have attained significance purely as a consequence of sample size.  

Our analysis here shows how these problems can be addressed in a distributional NHST approach, but only for simple counting experiments.  An important aim for future work is to extend this distributional NHST approach to other experimental designs (e.g. comparative experiments, factorial experiments).  For now, the best we can do is suggest caution in the interpretation of point-form NHST results from studies involving complex systems and studies involving large datasets.

 \newcommand{\noop}[1]{}

  \pagebreak  \section*{Appendix}
  
  Let $f( p;a,b)$  be the probability of $p_A$ having a value between $p$ and $p +\Delta p$, when $p_A \sim beta(a,b)$ (this is the normalised probability density function for the Beta distribution $ beta(a,b)$ as in Equation \ref{eq:beta}).  Let    
  \begin{equation}
  I(r; a,b) = \int_0^r f( p;a,b)\mathrm{d}p
  \end{equation}   
  be the probability of getting $p_A \leq r$ when $p_A  \sim beta(a,b)$ (this is the cumulative distribution function for this probability density, called the regularised incomplete Beta function).  In this appendix we give a proof that  for  $K < N/2$ and $a \geq 1$
  \begin{eqnarray} 
  \label{eq:betaBinomial_greater_beta}
  P(k \leq K |N, p \sim beta(a,a) )= I(K/N; a,a) +\epsilon 
  \end{eqnarray}
  where   $\epsilon$ is always positive and approaches $0$ as $N$ approaches infinity.
  
  By definition
  \begin{equation*}
  P(k \leq K |N, p_A \sim beta(a,a)) = \int_{p=0}^1 P(k \leq K |N, p) f(p;a,a )\mathrm{d}p
  \end{equation*}
  where $ P(k \leq K |N, p)$ is the cumulative Binomial function.
  Let  $p_k=K/N$ and we can write
  \begin{eqnarray*}
  	\int_{p=0}^{p_k}  P(k \leq K |N, p)  f(p;a,a )\mathrm{d}p &=& \int_{p=0}^{p_k} [1-P(k > K |N, p) ]  f(p;a,a)\mathrm{d}p \\
  	&=& \int_{p=0}^{p_k} f(p;a,a )\mathrm{d}p-\int_{p=0}^{p_k} P(k > K |N, p)  f(p;a,a )\mathrm{d}p \\	 
  	&= & I(K/N; a,a)-\int_{p=0}^{p_k} P(k >K |N, p)  f(p;a,a )\mathrm{d}p 
  \end{eqnarray*}
  and we have 
  \begin{eqnarray*}
  	P(k \leq rN |N, p_A \sim beta(a,a)) =    I(K/N; a,a) +E(K,N,a) 
  \end{eqnarray*}
  where
  \begin{eqnarray*}
  	E(K,N,a)= \int_{p={p_k}}^1 P(k \leq K |N, p)   f(p;a,a )\mathrm{d}p - \int_{p=0}^{p_k} P(k > K |N, p)   f(p;a,a )\mathrm{d}p  
  \end{eqnarray*}
  The variance of the cumulative Binomial functions $P(k \leq K |N, p)$ and $P(k > K |N, p)$ both fall to a limit of $0$ as $N$ approaches infinity.  This means that the probability of getting $k \leq K$ (when $p \geq p_k$) and the probability of getting $k >K$ (when $p \leq p_k$) both fall to a limit of $0$ as $N$ approaches infinity and so $E(K,N,a)$ approaches the limit of $0$ as $N$ approaches infinity.
  
  We wish to prove that $E(K,N,a) \geq 0$ for all $N$ and $a \geq 1$ where $K < N/2$.  The proof is inductive.  
  Since the probability density function $f(p,a,a)$ is symmetric we have
  \begin{eqnarray*} 
  	\int_{p=1-p_k}^{1} P(k \leq K |N, p)   f(p;a,a )\mathrm{d}p = \int_{p=0}^{p_k}  P(k \geq N- K |N, p) f(p;a,a )\mathrm{d}p  
  \end{eqnarray*}
  and we can rewrite $E(K,N,a)$ as
  \begin{eqnarray*} 
  	E(K,N,a)&=&  \int_{p={p_k}}^{1-p_k} P(k \leq K |N, p)   f(p;a,a )\mathrm{d}p - \int_{p=0}^{p_k} \left[ P(k > K |N, p) -P(k \geq N- K |N, p) \right] f(p;a,a )\mathrm{d}p  
  \end{eqnarray*}
  Since $K$ is an integer and since by assumption $K < N/2$ we have $K+1 \leq N-K$. When $K+1 \leq N-K$ we have
  \begin{eqnarray*}  
  	P(k > K |N, p) - P(k \geq N- K |N, p)  = P(K < k < N-K |N, p)
  \end{eqnarray*} 
  for all values of $p$.  
  From Equations \ref{eq:beta} and \ref{eq:beta_recurrence}  we have
  \begin{equation*}
  f(p;a+1,a+1)=    \frac{2(2a+1)}{a} p(1-p) f(p;a,a)
  \end{equation*}   
  and combining these results we have
  \begin{eqnarray*} 
  	E(K,N,a+1) &=& \frac{2(2a+1)}{a}  \int_{p={p_k}}^{1-p_k} P(k \leq K |N, p) \ p(1-p)  f(p;a,a )\mathrm{d}p \\ & & -\ \ \  \frac{2(2a+1)}{a}  \int_{p=0}^{p_k}  P(K < k < N-K |N, p) \ p(1-p) f(p;a,a )\mathrm{d}p  
  \end{eqnarray*}
  Since $p(1-p)$ is symmetrical and unimodal with its maximum at $p=0.5$ and minima at $p=0$ and $p=1$, we have
  \begin{eqnarray*} 
  	\int_{p={p_k}}^{1-p_k} P(k \leq K |N, p) \ p(1-p)  f(p;a,a )\mathrm{d}p &  \geq  & p_k(1-p_k) \int_{p={p_k}}^{1-p_k} P(k \leq K |N, p)   f(p;a,a )\mathrm{d}p
  \end{eqnarray*}        
  and similarly
  \begin{eqnarray*} 
  	\int_{p=0}^{p_k} P(K < k < N-K |N, p) \ p(1-p) f(p;a,a )\mathrm{d}p 
  	& \leq  &  p_k(1-p_k) \int_{p=0}^{p_k}P(K < k < N-K |N, p) f(p;a,a )\mathrm{d}p
  \end{eqnarray*} 
  and given these upper and lower bounds we get    
  \begin{eqnarray*}  
  	E(K,N,a+1) &\geq & p_k(1-p_k) \frac{2(2a+1)}{a} \\
  	&&\ \ \  \times \left[ \int_{p={p_k}}^{1-p_k} P(k \leq K |N, p)  f(p;a,a )\mathrm{d}p- \int_{p=0}^{p_k}  P(K < k < N-K |N, p)  f(p;a,a )\mathrm{d}p  \right] \\
  	&\geq & p_k(1-p_k) \frac{2(2a+1)}{a} E(K,N,a)  
  \end{eqnarray*}
  Since $p_k$,$1-p_k$ and $a$ are all necessarily non-negative we see that  if $E(K,N,a) \geq 0$ (and $K < N/2$) then $E(K,N,a+1) \geq 0$ necessarily holds.   This completes the inductive step.   
  
  With $a=1$ we have $ p_A \sim beta(1,1)$ and $p_A$ is distributed uniformly in the range $0\ldots 1$.  With $p_A \sim U(0,1)$ each possible value of $k$ from 0 to $N$ is equally likely to occur, and since there are $N+1$ such values in total, of which $K+1$ are less than or equal to $K$, the probability of getting $k \leq K$ is  
  \begin{equation*} 
  P(k \leq K| N, p \sim beta(1,1)) = \frac{ K+1}{N+1} \geq \frac{ K}{N}
  \end{equation*} 
  But with $ p_A \sim beta(1,1)$ we have $I(K/N; 1,1)=K/N$, and so 
  \begin{equation*} 
  P(k \leq K | N, p \sim beta(1,1))  \geq I(K/N; 1,1) 
  \end{equation*} 
  and $E(K,N,a) \geq 0$ holds for all $N$ and $K$ when $a=1$.   By our inductive step this means that  $E(K,N,a) \geq 0$  holds for all $a \geq 1$ and $K < N/2$ giving our required result.

 \end{document}